\documentclass[aps,prc,twocolumn,floatfix,superscriptaddress,nofootinbib]{revtex4-2}

\usepackage{graphicx}
\usepackage[urlcolor=green]{hyperref}
\usepackage{amsmath}
\usepackage{physics}

\newcommand{\Kompost}{K{\o}MP{\o}ST}

\begin{document}

\author{Nicolas Borghini} \email{borghini@physik.uni-bielefeld.de}
\affiliation{Fakult\"at f\"ur Physik, Universit\"at Bielefeld, D-33615 Bielefeld, Germany}
\author{Renata Krupczak} \email{rkrupczak@physik.uni-bielefeld.de}
\affiliation{Fakult\"at f\"ur Physik, Universit\"at Bielefeld, D-33615 Bielefeld, Germany}
\author{Hendrik Roch} \email{Hendrik.Roch@wayne.edu}
\affiliation{Department of Physics and Astronomy, Wayne State University, Detroit, Michigan 48201, USA}

\title{Comparing matching prescriptions between pre-equilibrium and hydrodynamic models in high-energy nuclear collisions}

\begin{abstract}
State-of-the-art simulations of high-energy nuclear collisions rely on hybrid setups, involving in particular a pre-equilibrium stage to let the system evolve from a far-from-equilibrium initial condition towards a near-equilibrated state after which fluid dynamics can be applied meaningfully. 
A known issue is the mismatch between the equation of state in the fluid-dynamical evolution and the effective one in the previous stage, which leads to discontinuities at the interface between the two models.
Here we introduce a new matching prescription at this interface, based on the entropy, and we compare it with the standard one relying on local energy conservation. 
We study the behavior of various quantities at the switching time between the models and investigate a number of final-state hadronic observables.
For the latter, we show that they are not modified significantly by the choice of matching prescription, provided an appropriate normalization is chosen for the initial state. 
In turn, our approach reduces sizeably the ratio of bulk over thermodynamic pressure at the beginning of the fluid-dynamical stage.
\end{abstract}

\maketitle

\section{Introduction}
\label{s:intro}

Collisions of heavy nuclei at high energies create a hot and dense strongly-interacting system~\cite{Busza:2018rrf, Elfner:2022iae}, whose bulk evolution can satisfactorily be described by relativistic dissipative fluid dynamics~\cite{Teaney:2009qa} with the equation of state determined in finite-temperature quantum chromodynamics (QCD) computations on the lattice~\cite{Borsanyi:2010cj, Borsanyi:2013bia, HotQCD:2014kol}.

These fluid-dynamical simulations are complemented by further models.
First, initial conditions are needed to describe the direct outcome of the initial nucleus-nucleus collision, for which there exist parametric models~\cite{Moreland:2014oya, Ke:2016jrd} as well as approaches more grounded on the saturation regime of QCD~\cite{Schenke:2012wb, Paatelainen:2012at, Garcia-Montero:2023gex}. 
Then, since the system thus created is far from thermodynamic equilibrium, so that (isotropic) fluid dynamics may not be applied meaningfully, pre-equilibrium models are often employed to describe the hydrodynamization~\cite{Schlichting:2019abc}, and in particular to reduce the large gradients present in the initial state~\cite{NunesdaSilva:2020bfs}.
Eventually, after the end of the hydrodynamic evolution, an ``afterburner''~\cite{SMASH:2016zqf} models the interactions of hadrons that become the relevant degrees of freedom when the system description as a fluid is no longer a good approximation.
For a faithful physical modeling, it is crucial to investigate the best way to match two successive approaches in such hybrid setups. 

In this study, we focus on the interface between the pre-equilibrium and fluid-dynamical stages. 
For the early evolution, there are essentially two classes of models: 
either free streaming, in which classically the degrees of freedom propagate at the speed of light for a short amount of time without interaction~\cite{Liu:2015nwa}.
Or one lets the QCD degrees of freedom interact, relying on an effective kinetic theory description~\cite{Kurkela:2018vqr,Liyanage:2022nua}. 
Both approaches assume massless degrees of freedom, so that when the system is (almost) equilibrated the equation of state is the conformal one, $e = 3P$.
This is at variance with the equation of state used in the hydrodynamic stage, which realistically should be non-conformal.
As a result, some of the fluid-dynamical quantities are discontinuous when switching from one model to the other. 

An approach to handle this issue is to assume that the free-streaming velocity $v$ is smaller than the speed of light; $v$ is then treated as a fit parameter~\cite{Nijs:2020roc, daSilva:2022xwu}.
However, because of the fluctuations in the initial state --- not only event-by-event but also point-to-point in a given event ---, such a recipe with a position independent $v$ can only mitigate the mismatch on average, but cannot hold equally well at each point in the final state of the pre-equilibrium evolution.

In the present study, we introduce a new matching prescription between the pre-equilibrium and fluid-dynamical models, which aims at addressing some of the issues that arise at this interface. 
To test this approach, we perform hybrid numerical simulations, whose setup, together with our proposed matching recipe, is detailed in Sec.~\ref{s:setup}. 
We compare simulations with the new approach to computations with the standard one, focusing first on the switching time between pre-equilibrium and fluid dynamics, then on a few final-state hadronic observables (Sec.~\ref{s:results}). 
Our findings are summarized and discussed in Sec.~\ref{s:discussion}, in which we also mention future perspectives.

\section{Model}
\label{s:setup}

In this study, we simulate Pb-Pb collisions at 5.02~TeV (without conserved charges) using a hybrid setup whose successive steps we now detail~\cite{ourhybrid}. 

\subsection{Hybrid approach}

\noindent {\bf 1.} The initial states are generated with a Monte Carlo (MC) Glauber model that determines a transverse profile from the positions of participant nucleons and their binary collisions~\cite{Miller:2007ri, dEnterria:2020dwq}.
The precise parameters of the model are the same as in Ref.~\cite{Borghini:2022iym}: since they are not crucial for the present study, we do not repeat them here, but only mention the Gaussian-smearing radius of 0.4~fm used in the energy deposition, which gives the typical scale of ``hot spots'' in the initial state. 
The only difference with the setup of Ref.~\cite{Borghini:2022iym} is that we do not fix the impact parameter value $b$, but sample it randomly with a probability $p(b)\,\dd{b}\propto b\,\dd{b}$.
Each MC Glauber transverse profile, scaled by a multiplicative $K$-factor which is set such that it leads to the correct final-state charged multiplicity (see below), is then interpreted as an energy-density distribution $e_0(\vb{x})$ at midrapidity, multiplied by the starting time $\tau_0$ of the dynamical evolution.

\noindent {\bf 2.} The profiles from the previous step are used as initial condition, at a fixed proper time $\tau_0=0.2$~fm,\footnote{Throughout the paper we use $c=\hbar=k_B=1$.} in \Kompost~\cite{Kurkela:2018vqr}, which we use to model the dynamics in the pre-equilibrium stage.
Since the input consists of an energy density profile, \Kompost\ is run in the mode with only energy perturbations, to propagate the inhomogeneities in $e_0(\vb{x})$.  
The specific shear viscosity in \Kompost\ is taken to be $\eta/s = 0.16$ and the effective number of degrees of freedom to be $\nu_{\text{eff}} = 40$.
The hydrodynamization time $\tau_{\rm hydro}$ at which \Kompost\ ends is set equal to 1~fm. 
\Kompost\ then provides at each point of the transverse plane an energy-momentum tensor, which is decomposed using the Landau matching condition into a local energy density $e_{\rm K}(\vb{x})$, velocity $u^\mu(\vb{x})$ and shear stress tensor $\pi^{\mu\nu}(\vb{x})$. 
Since \Kompost\ is conformal, the ``thermodynamic'' pressure is $P_{\rm K}(\vb{x})=e_{\rm K}(\vb{x})/3$, while the bulk pressure $\Pi_{\rm K}(\vb{x})$ vanishes.

\noindent {\bf 3.} At $\tau_{\rm hydro}$ the fluid-dynamical stage begins, which we model with the relativistic dissipative hydrodynamics code MUSIC~\cite{Schenke:2010rr, Schenke:2010nt, Paquet:2015lta} in its boost-invariant mode, and using the lattice QCD (LQCD) equation of state from the HotQCD Collaboration~\cite{HotQCD:2014kol}, matched at low temperatures with the hadron resonance gas used in SMASH.
How the energy density $e_{\rm K}(\vb{x})$ from \Kompost\ is used for the initial state of MUSIC is precisely the main difference between the two scenarios tested in this paper and will be specified in Sec.~\ref{ss:matching}. 
In turn, the velocity and shear stress tensor from \Kompost\ are used as is to initialize the energy-momentum tensor in MUSIC.
For the subsequent hydrodynamic evolution, we use a constant shear viscosity over entropy density ratio $\eta/s = 0.16$, as in \Kompost. 
For the bulk viscosity, we choose the temperature-dependent parameterization introduced in Ref.~\cite{Denicol:2009am} and whose impact on hadronic observables was studied in Ref.~\cite{Ryu:2015vwa, Ryu:2017qzn}. 
When a MUSIC cell reaches a temperature of 155~MeV, particlization takes place and the fluid cell is converted using the iSS package~\cite{Shen:2014vra} into a sample of hadrons.

\noindent {\bf 4.} The latter are fed into the hadronic transport approach SMASH~\cite{SMASH:2016zqf}, which models their further rescatterings and possible decays.
From the final state of SMASH we can then compute hadronic observables of interest, which we shall detail in Sec.~\ref{ss:observables}.

Using this overall hybrid setup --- with both matching prescriptions between \Kompost\ and MUSIC that will be discussed hereafter ---, we generated $2^{15}$ initial states with random impact parameter. 
Using the semi-empirical formula for final charged-hadron multiplicity based on the initial energy density introduced in Ref.~\cite{Giacalone:2019ldn}, with at first a guess for the $K$-factor at the end of the MC Glauber, we estimated the centrality classes in which the $2^{15}$ events are distributed. 
The precise value of the $K$-factor was then determined by fitting the respective centrality-dependent multiplicities to ALICE data~\cite{ALICE:2015juo}.

In each centrality bin, we also computed an average initial state, which we then let evolve dynamically with \Kompost+MUSIC+SMASH like one of the MC Glauber initial states. 
For the final-state observables presented in Sec.~\ref{ss:observables}, we actually computed 5000 random events with the full hybrid setup (for both matching prescriptions) in each centrality class.

\subsection{Matching the pre-equilibrium model to fluid dynamics}
\label{ss:matching}

As stated in the introduction, the mismatch between the conformal relation $e = 3P$ at the end of the pre-equilibrium stage and the LQCD equation of state unavoidably induces jumps in some fluid-dynamical quantities when one switches from \Kompost\ to MUSIC. 
In our simulations we compare two matching prescriptions, with a focus on different thermodynamic quantities. 

On the one hand, we use the by now standard recipe, consistent with Landau matching, in which the energy density $e_{\rm K}(\vb{x})$ at a given transverse position (in practice, in a cell) at the end of the pre-equilibrium model, here \Kompost, becomes the energy density $e_{\rm M}(\vb{x})$ at the same point in the hydrodynamic code, namely MUSIC: 
$e_{\rm M}(\vb{x}) = e_{\rm K}(\vb{x})$.
In the following, this approach will be referred to as ``energy matching''. 
Since we also feed the final velocity of \Kompost\ as is into MUSIC, this mode ensures local energy conservation. 
Eventually, the bulk pressure in MUSIC is initialized as 
\begin{equation}
\label{eq:Pi_M_e-matching}
\Pi_{\rm M}(\vb{x}) \equiv 
\frac{e_{\rm K}(\vb{x})}{3} - P_{\rm LQCD}\big(e_{\rm K}(\vb{x})\big), 
\end{equation}
which also equals $P_{\rm K}(\vb{x}) - P_{\rm LQCD}\big(e_{\rm K}(\vb{x})\big)$, such that the sum $P+\Pi$ is continuous at the model interface. 

On the other hand, our proposed ``entropy matching'' prescription goes as follows. 
From the energy density $e_{\rm K}(\vb{x})$ in a cell in \Kompost\ at $\tau_{\rm hydro}$, the effective number of freedom $\nu_{\text{eff}}$, and the conformal equation of state $e = 3P$, we compute the equivalent equilibrium entropy density $s_{\rm K,eq.}(\vb{x})$:
\begin{equation}
\label{eq:s_K}
s_{\rm K,eq}(\vb{x}) \equiv 
\frac{4}{3}\bigg(\frac{\nu_{\text{eff}}\pi^2}{30} \bigg)^{\!\!1/4} e_{\rm K}(\vb{x})^{3/4}.
\end{equation}
The ``matching'' consists in postulating that this becomes the initial local equilibrium entropy density $s_{\rm M,eq.}(\vb{x})$ in MUSIC: $s_{\rm M,eq.}(\vb{x})=s_{\rm K,eq.}(\vb{x})$. 
We then determine which energy density $e_{\rm LQCD}$ corresponds to this equilibrium entropy density using the LQCD equation of state: this defines the energy density $e_{\rm M}(\vb{x}) \equiv e_{\rm LQCD}\big(s_{\rm M,eq.}(\vb{x})\big)$ with which we initialize the components of the energy-momentum tensor in MUSIC.\footnote{That is, we really use an energy density as input in MUSIC, not an entropy density, even though our $e_{\rm M}(\vb{x})$ is obtained via the corresponding $s_{\rm M,eq.}(\vb{x})$.}

Obviously, this prescription leads to non-conservation of energy, both locally and in general globally, at the interface between the pre-equilibrium and fluid-dynamical descriptions. 
As we shall discuss below, the recipe $s_{\rm M,eq.}(\vb{x})=s_{\rm K,eq.}(\vb{x})$ does not ensure entropy conservation either, since we only match the (equivalent) equilibrium entropy densities. 
However, we can still ensure local momentum conservation, as is also done in the energy-matching prescription. 
For that, since we feed the velocity $u^\mu$ and shear-stress tensor $\pi^{\mu\nu}$ from \Kompost\ into MUSIC, it is sufficient to keep the total pressure $P+\Pi$ constant over the switch. 
Using $P_{\rm K} = e_{\rm K}/3$ and $\Pi_{\rm K}=0$ in \Kompost, while the thermodynamic pressure in MUSIC is related to the energy density via the LQCD equation of state, we thus determine the initial bulk pressure $\Pi_{\rm M}(\vb{x})$ in MUSIC by
\begin{equation}
\label{eq:Pi_M_s-matching}
\Pi_{\rm M}(\vb{x}) \equiv \frac{e_{\rm K}(\vb{x})}{3} - P_{\rm LQCD}\big(e_{\rm M}(\vb{x})\big). 
\end{equation}
Since $e_{\rm M}(\vb{x})\neq e_{\rm K}(\vb{x})$, this differs from Eq.~\eqref{eq:Pi_M_e-matching} in the energy-matching prescription.

Before we present the results of our simulations, let us comment on the new matching prescription we propose. 
Since the flow velocity $u^\mu(\vb{x})$ at the end of \Kompost\ is also fed in unchanged into MUSIC, our recipe ensures the continuity of the product $s_{\rm eq}(\vb{x})u^\mu(\vb{x})$ at the interface. 
In the absence of conserved charge as we assume here, this product coincides in the relativistic fluid with  the entropy current $S^\mu(\vb{x})$ up to terms of second order in gradients~\cite{Israel:1976tn,Israel:1979wp}, since $u^\mu(\vb{x})$ is such that the associated local rest frame is the Landau frame. 
Assuming, as is implicitly done, that the system at the end of the \Kompost\ evolution is close to equilibrium, then $s_{\rm eq,K}(\vb{x})u^\mu(\vb{x})$ is also the entropy current up to second-order terms. 
Yet even though the shear stress tensor is taken to be continuous at the switch between both models, the non-equilibrium corrections to $S^\mu(\vb{x})$ they induce do not automatically match, because they depend on the (equivalent) temperature, which is not the same in \Kompost\ and MUSIC at a given $s_{\rm eq}$ due to the different equations of state. 
Also, any correction to the entropy current involving the bulk pressure is zero in \Kompost\ but generally non-zero in MUSIC. 
Thus, while we are aware that a better-motivated prescription would be to ensure the continuity of the total entropy current, yet we were not able to do it in this exploratory study. 
In addition, the correct approach would be to compute $S^\mu(\vb{x})$ in \Kompost\ without assuming that it describes an almost equilibrated dissipative fluid, but we do not know how this can be done, unless an underlying kinetic model is assumed. 
Yet our recipe in the present paper ensures approximate entropy conservation at the interface between pre-equilibrium and fluid dynamics, while it is not conserved in the energy-matching prescription~\cite{Liu:2015nwa}.

\section{Results}
\label{s:results}

\begin{figure*}[ht!]
\includegraphics[width=0.49\linewidth]{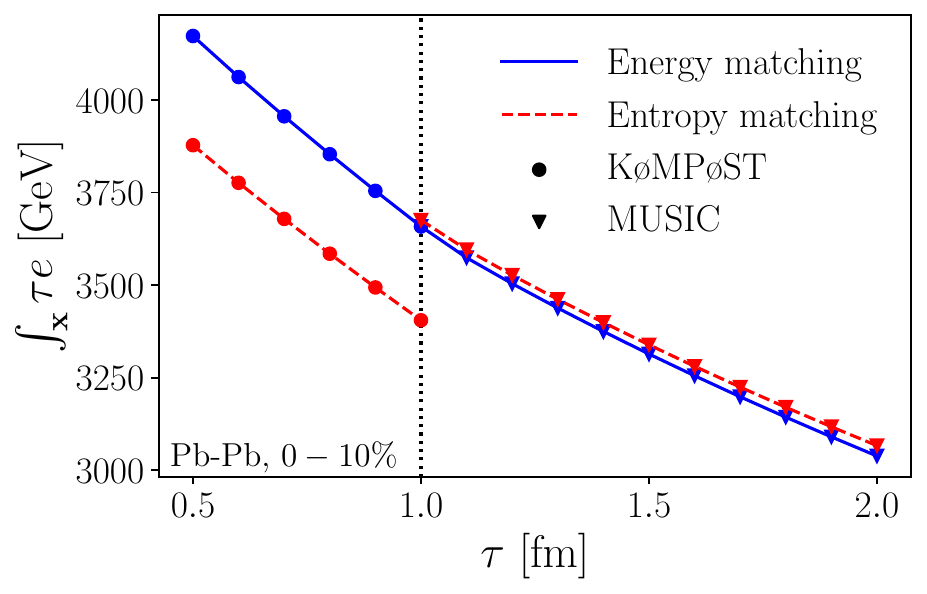}
\includegraphics[width=0.49\linewidth]{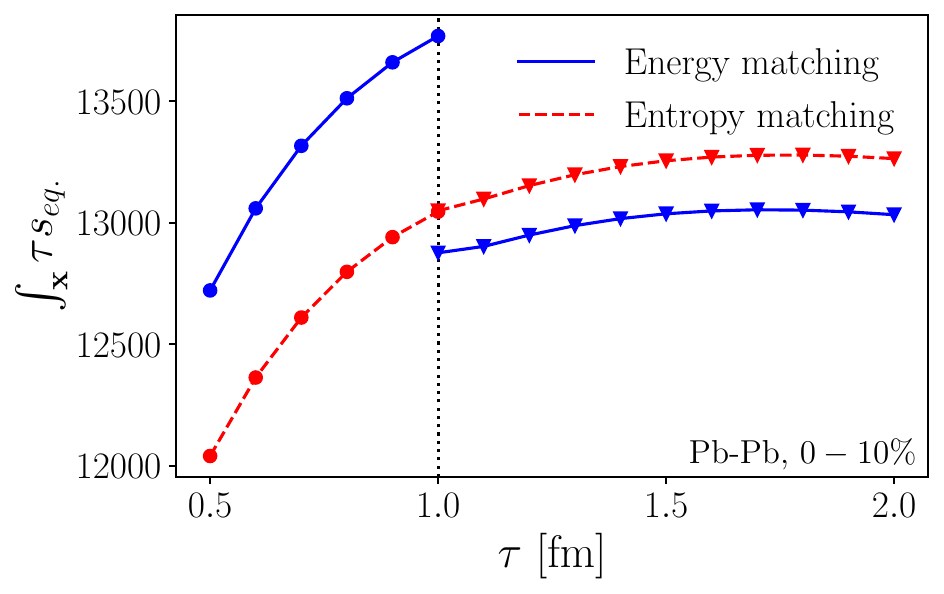}
\includegraphics[width=0.49\linewidth]{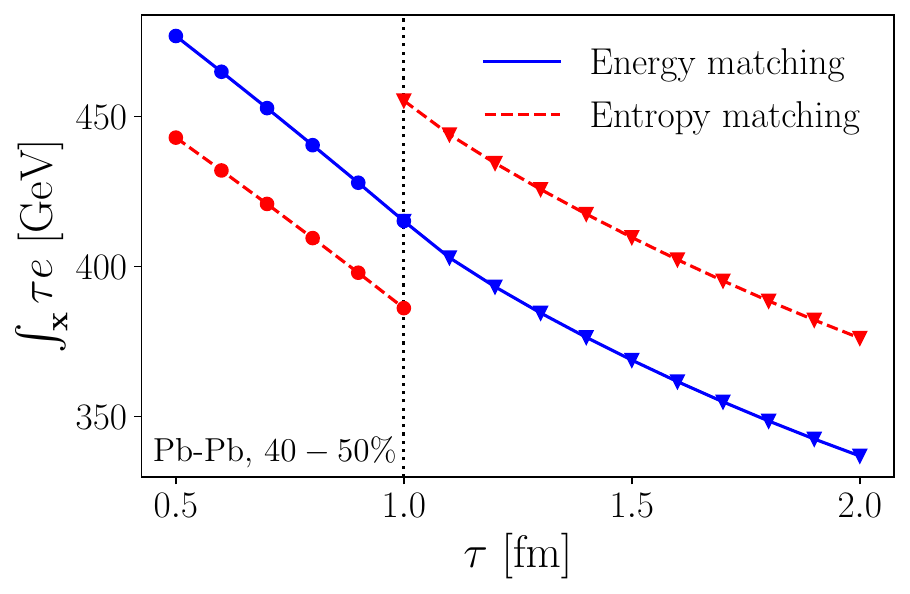}
\includegraphics[width=0.49\linewidth]{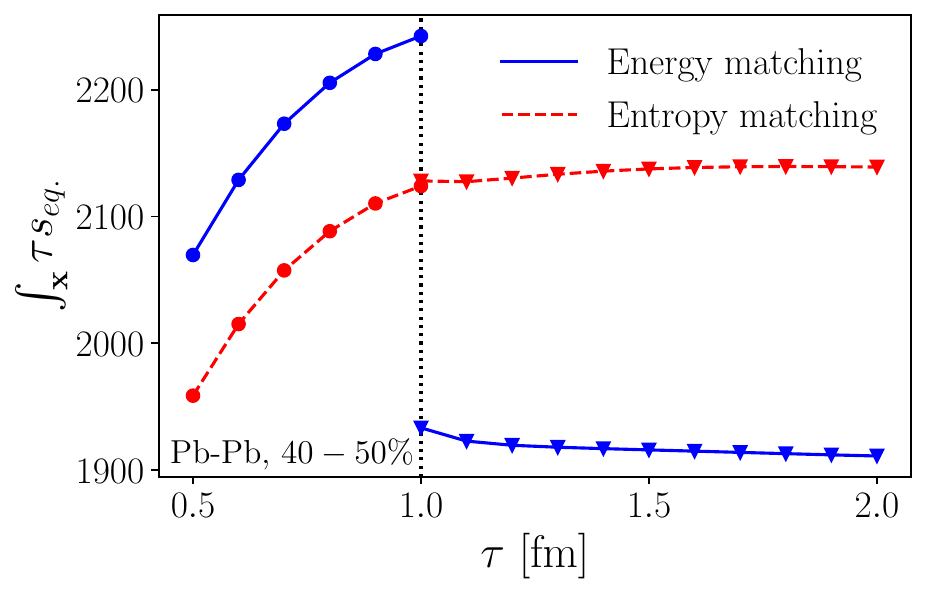}\vspace{-3mm}
\caption{Time dependence of the energy (left) and (equivalent) equilibrium entropy (right) per unit rapidity, using the energy-matching (blue full line) and entropy-matching (red dashed line) prescriptions. 
The vertical dotted line stands for the switching time $\tau_{\rm hydro}$ between \Kompost\ and MUSIC. 
The top plots are for the average central (0--10\%) event and the bottom ones are for the average event in the 40--50\% centrality class.}
\label{fig:e_and_s}
\end{figure*}

In this section we present our results, starting with the behavior of several quantities around the transition from \Kompost\ to MUSIC (Sec.~\ref{ss:comparisons}), before going to final-state observables (Sec.~\ref{ss:observables}).

\subsection{Thermodynamic and fluid-dynamical quantities around the switching time}
\label{ss:comparisons}

\begin{figure*}[ht!]
\includegraphics[width=0.49\linewidth]{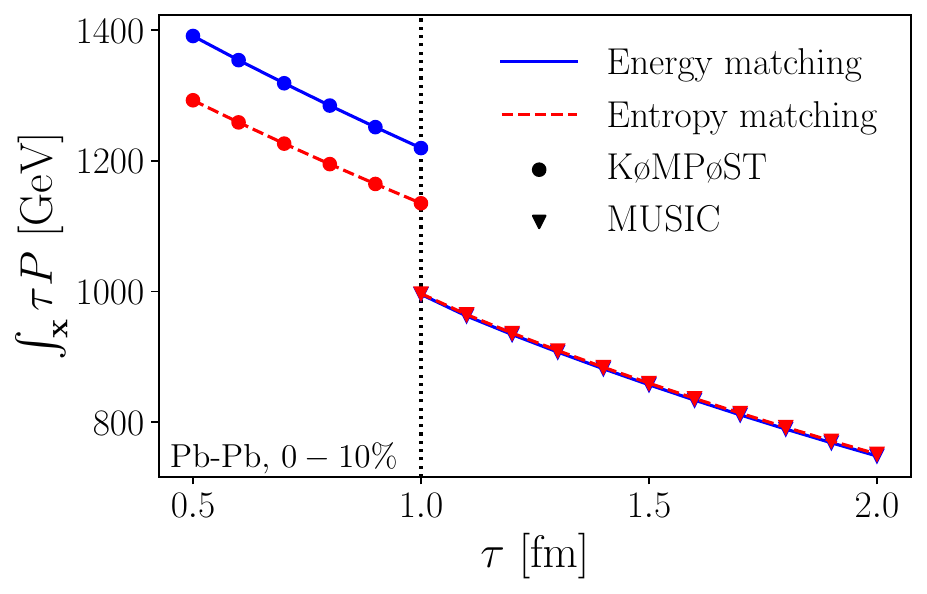}
\includegraphics[width=0.49\linewidth]{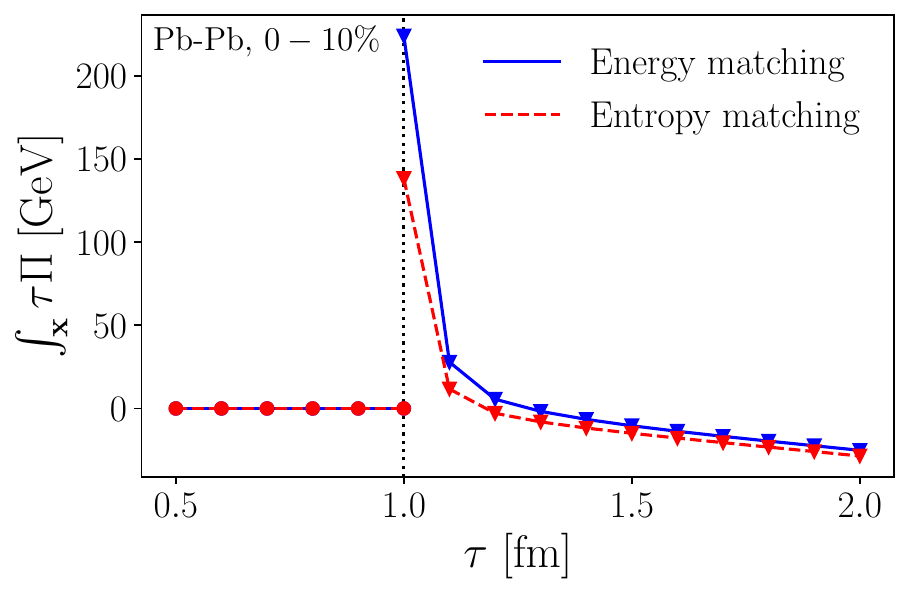}
\includegraphics[width=0.49\linewidth]{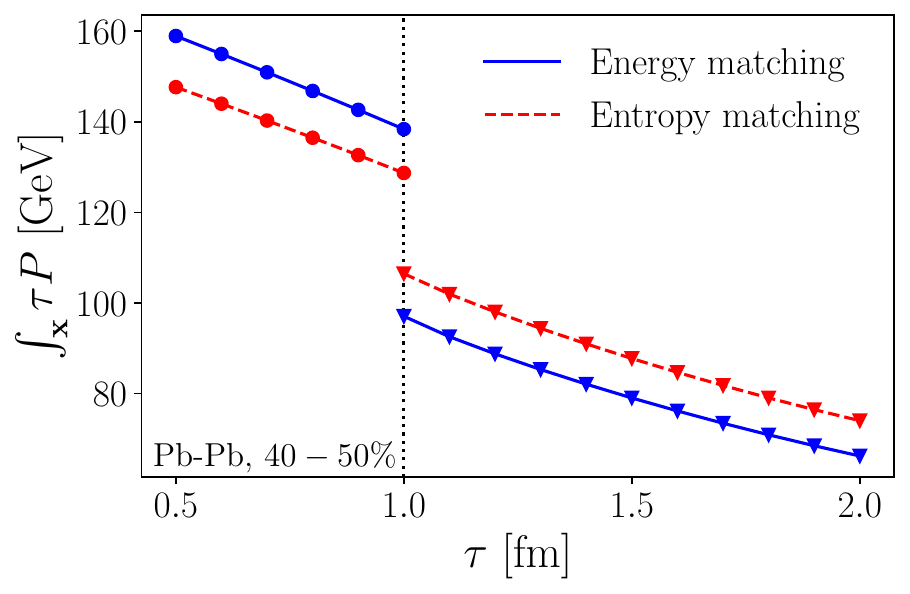}
\includegraphics[width=0.49\linewidth]{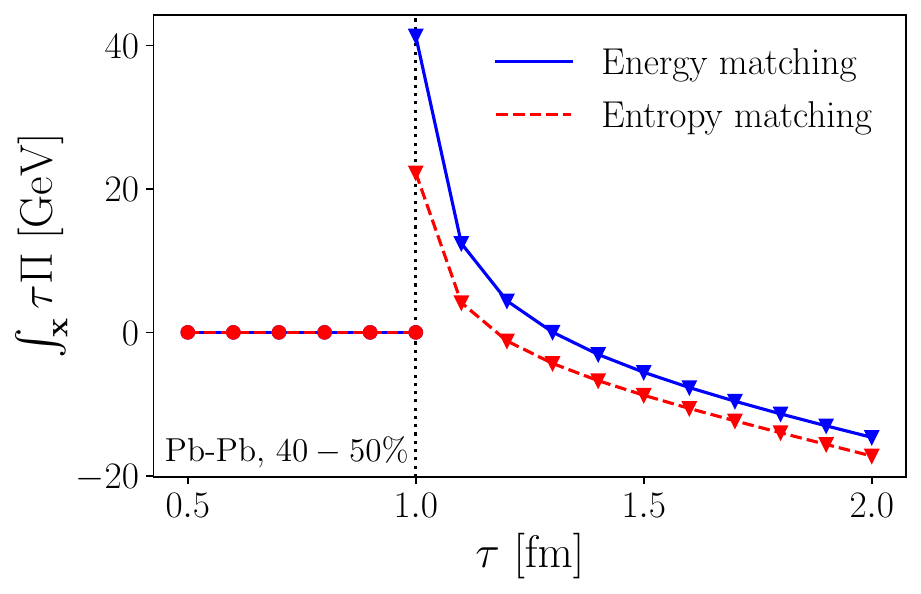}\vspace{-3mm}
\caption{Time dependence of the thermodynamic pressure (left) and bulk pressure (right) per unit rapidity, using the energy-matching (blue, full) and entropy-matching (red, dashed) prescriptions, for the average central (0--10\%, top) and peripheral (40--50\%, bottom) events.
The vertical dotted line stands for the switching time $\tau_{\rm hydro}$ between \Kompost\ and MUSIC.}
\label{fig:p_and_Pi}
\end{figure*}

As mentioned in the previous section, the multiplicative factor needed to turn the output of the MC Glauber into a transverse energy density profile at $\tau_0$ is determined such that the final charged multiplicity fits the experimental data. 
Since the two matching prescriptions induce different discontinuities, which we shall presently look into, at the switch from \Kompost\ to MUSIC, they require different factors, namely $K \simeq 51.43$~GeV/fm$^2$ for energy matching and the somewhat smaller value $K\simeq 47.73$~GeV/fm$^2$ for entropy matching. 

In Figs.~\ref{fig:e_and_s} and \ref{fig:p_and_Pi} we display the time dependence between 0.5 and 2~fm, i.e.\ around the hydrodynamization time $\tau_{\rm hydro}$, of several quantities integrated over the whole transverse plane and within one unit of spatial rapidity. 
In both figures, we use the ``average events'' resulting from the dynamical evolution of the average initial states in the 0--10\% (top panels) and 40--50\% (bottom panels) centrality classes. 
We checked that these average events are representative of the findings in individual, ``random'' events in the corresponding centrality classes. 

Figure~\ref{fig:e_and_s} shows the evolution of energy (left) and equivalent equilibrium entropy (right), i.e.\ the quantities whose continuity at the local level determines our two matching prescriptions. 
Accordingly, the energy is continuous at $\tau_{\rm hydro}$ when using energy matching, while the integral of $s_{\rm eq.}$ is continuous within the entropy-matching approach. 
We recall that $s_{\rm eq.}(\vb{x})$ is defined by Eq.~\eqref{eq:s_K} in the pre-equilibrium stage, while it is given by LQCD thermodynamics in MUSIC.

In contrast, the mismatch between the equations of state in \Kompost\ and MUSIC means that there are discontinuities at the switching time. 
Thus, energy is discontinuous when applying entropy matching (left, red curves), while reciprocally the integral of $s_{\rm eq.}$ is discontinuous in the simulation with energy matching (right, blue curves). 
Due to the different initial $K$-factors, the values of the energy --- and thus the equivalent equilibrium entropy ---  necessarily differ in the two matching prescriptions in the pre-equilibrium stage. 
Yet for those quantities with a jump at $\tau_{\rm hydro}$, the values may become closer in MUSIC, because the final charged multiplicity should be the same.
This trend is not so clear in the peripheral events, but one should pay attention that the difference between the values in MUSIC after using the two prescriptions are actually of the same absolute size (about 30~GeV for energy) as in central events. 
Furthermore, one should keep in mind that the integral of the equilibrium entropy $s_{\rm eq.}$ is not the total entropy (per unit rapidity) in the (MUSIC) fluid. 
In addition to $s_{\rm eq.}$, one should also consider the non-equilibrium contributions from shear stresses and bulk pressure~\cite{Israel:1976tn,Israel:1979wp} --- and as we shall discuss next, bulk pressure in MUSIC does depend on the matching recipe. 
This also explains why the slight decrease of the integral of $s_{\rm eq.}$ in the MUSIC stage is not in contradiction with the second law of thermodynamics. 
In fact, there are also parts of the system that have already particlized, and which we do not take into account in those plots. 

In Fig.~\ref{fig:p_and_Pi}, we show the evolution of thermodynamic (left) and bulk pressure (right) integrated over the transverse plane. 
We recall that in \Kompost\ $P=e/3$ and $\Pi=0$, so that the portions of these plots before $\tau_{\rm hydro}$ are either degenerate with Fig.~\ref{fig:e_and_s} or trivial. 
Indeed, the most interesting is the behavior at $\tau_{\rm hydro}$, where these pressures are discontinuous. 
In all cases, the jumps in $P$ or $\Pi$ are smaller with the entropy-matching prescription than with energy matching. 
When equating $s_{\rm eq.}$ in MUSIC and \Kompost, the energy density at the start of MUSIC is larger than at the end of \Kompost: 
this corresponds to a higher temperature, so that the LQCD equation of state is closer to being conformal, as it is in \Kompost, which explains why the gap in thermodynamic pressure is smaller. 
In turn, since the sum $P+\Pi$ is continuous this results in an equally smaller gap in bulk pressure, such that the initial $\Pi$ in MUSIC is smaller with entropy matching. 

\begin{figure}
\includegraphics[width=\linewidth]{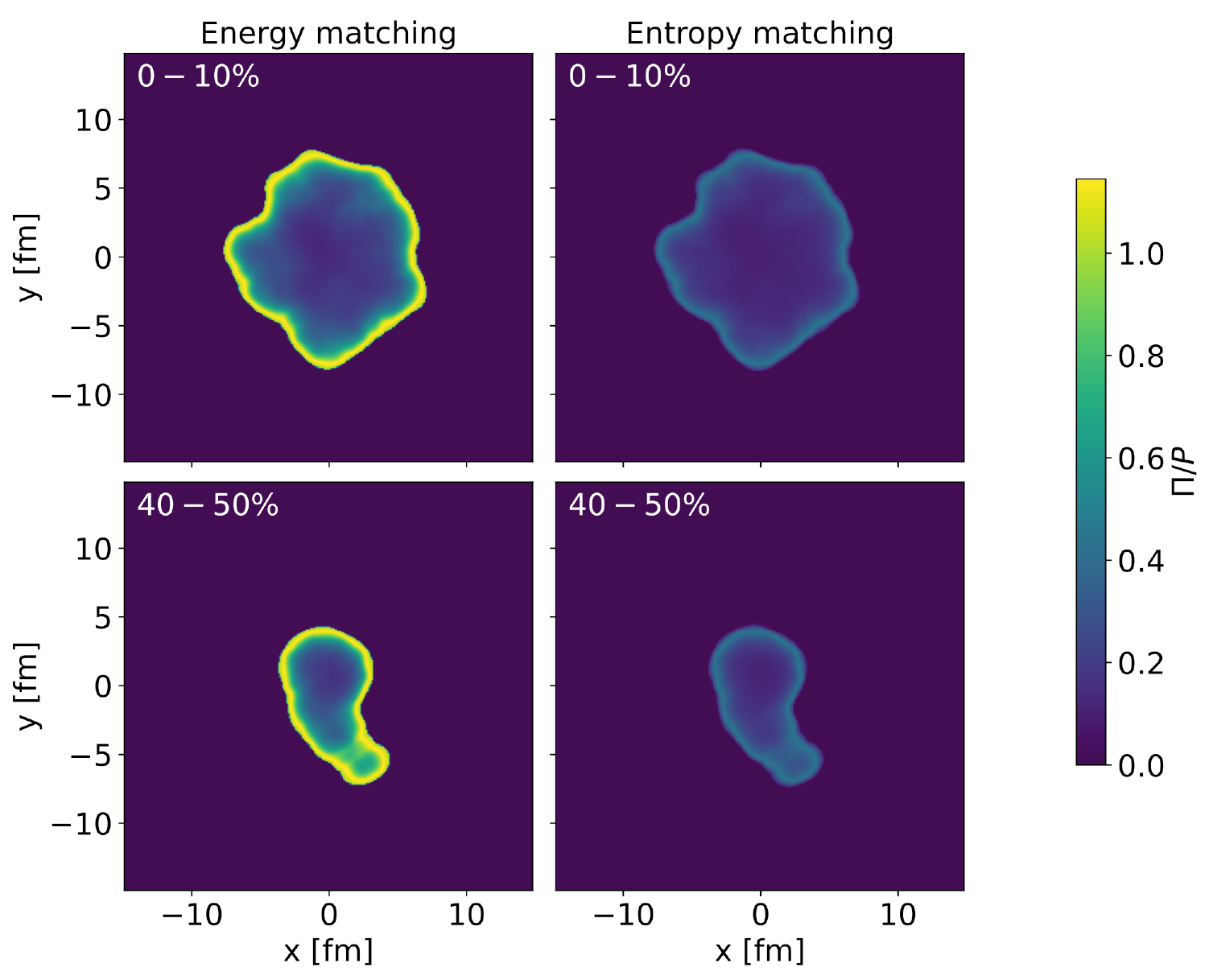}
\caption{Transverse-plane dependence of the ratio $\Pi/P$ between the bulk and thermodynamic pressures at the initial time $\tau_{\rm hydro}=1$~fm of the fluid-dynamical description, for a random central (0--10\%, top panels) and peripheral (40--50\%, bottom) event, using energy (left) and entropy matching (right).}\vspace{-2mm}
\label{fig:Pi/P}
\end{figure}

This reduction of the bulk pressure is actually an important positive feature of the entropy-matching prescription. 
In Fig.~\ref{fig:Pi/P} we plot the ratio of bulk over thermodynamic pressure $\Pi/P$ across the transverse plane in the initial state of MUSIC at $\tau_{\rm hydro}$, for random central (top panels) and peripheral (bottom) events with either energy (left) or entropy (right) matching prescription, where we show only the regions of the system with a temperature larger than 155~MeV. 
One sees that $\Pi/P$ is significantly reduced when using $s_{\rm eq}$-matching: it is at most about 0.5, while the standard energy-matching recipe leads to the presence of fluid cells with $\Pi/P > 1$, in particular towards the edges of the fireball~\cite{NunesdaSilva:2020bfs,daSilva:2022xwu}. 
Suppressing these regions with $\Pi/P \sim 1$ is important since in that regime the applicability of dissipative fluid dynamics is questionable.\footnote{Since the bulk pressure $\Pi$ resulting from the matching at $\tau_{\rm hydro}$ is positive, $|\Pi/P| > 1$ does not signal the possible onset of cavitation, as may happen with a negative $\Pi$ around the QCD crossover region~\cite{Rajagopal:2009yw,Denicol:2015bpa,Byres:2019xld,Chiu:2021muk}.}
In that respect, the new proposed prescription makes the fluid-dynamical evolution more reliable.\footnote{Note that in all simulations in the present study, the normal regulators in MUSIC are used, to avoid the instability in hydrodynamics. This is one reason why the bulk pressure decreases fast in Fig.~\ref{fig:p_and_Pi}.}
Besides, several studies~\cite{NunesdaSilva:2020bfs, Liu:2015nwa} have shown that too large a bulk pressure affects some final observables like the flow coefficients $v_n$ and the average transverse momentum $\expval{[p_T]}$. 

It is possible to investigate the behavior at $\tau_{\rm hydro}$ of further physical quantities, which are always consequences of the jumps in energy or equilibrium entropy. 
For instance, enthalpy is discontinuous with both prescriptions, being reduced for energy matching and increasing for entropy matching. 
If one defines the spatial eccentricities with an energy-density weight, then there is a small jump in their evolution when the switch from pre-equilibrium to fluid dynamics is done with entropy matching. 
However, this does not affect the (mostly linear) correlation between the initial eccentricity and the final anisotropic flow coefficient.

\subsection{Final-state observables}
\label{ss:observables}

As already mentioned, the final charged-hadron multiplicity was fixed to determine the initial normalization for each approach, so that $N_{\rm ch.}$ is by construction almost the same with both matching prescriptions. 
We also computed a number of other observables in the final state of the SMASH evolution,\footnote{For that analysis, we used the SPARKX package~\cite{SPARKX1.2.1}.} to test how much they are affected by the change in matching recipe. 
The reported observables are integrated over all particles at mid-(pseudo)rapidity, $|\eta| < 0.5$, with no cut on transverse momentum.
More precisely, we looked at the average over events of the first five anisotropic-flow harmonics $v_n$ and of the average transverse momentum $[p_T]$ of charged hadrons. 
We also looked at an observable that is more sensitive to event-by-event fluctuations, namely the linear Pearson coefficient $\rho([p_T], v_2^2)$ for the correlation between $[p_T]$ and $v_2^2$~\cite{Bozek:2016yoj}.
For the flow coefficients we computed the modulus of the average of ${\rm e}^{{\rm i}n\varphi}$ over the charged hadrons in an event, where $\varphi$ is the particle azimuthal angle with respect to the known impact-parameter direction.

\begin{table}[t!]
\caption{\label{tab:1}Final-state observables at mid-rapidity for 5000 events in the centrality class 0--10\%.}
\begin{tabular}{c|c|c}
\toprule
Observable & Energy matching & Entropy matching \\
\hline
$\expval{N_{\rm ch.}}$ & $1808$ & $1828$ \\
$\expval{v_1}$ (\%) & $0.912$ & $0.883$ \\ 
$\expval{v_2} (\%)$ & $2.652$ & $2.597$ \\
$\expval{v_3}$ (\%) & $1.535$ & $1.530$ \\
$\expval{v_4}$ (\%) & $0.835$ & $0.830$ \\
$\expval{v_5}$ (\%) & $0.506$ & $0.514$ \\
$\expval{[p_T]}$ (MeV) & $702.45$ & $697.57$ \\
$\rho([p_T], v_2^2)$ & $0.0911$ & $0.1006$ \\
\hline
\end{tabular}
\end{table}

The values of these observables for central collisions are shown in Tab.~\ref{tab:1} for central events. 
One sees that the values of the observables are very similar with both matching prescriptions, mostly at the percent level.
The only exceptions are $v_1$, which varies by about 3\%, and especially $\rho([p_T], v_2^2)$, which changes by almost 10\%. 

\begin{table}[t!]
\caption{\label{tab:2}Final-state observables at mid-rapidity for 5000 events in the centrality class 40--50\%.}
\begin{tabular}{c|c|c}
\toprule
Observable & Energy matching & Entropy matching \\
\hline
$\expval{N_{\rm ch.}}$ & $259$ & $282$ \\
$\expval{v_1}$ (\%) & 1.476 & 1.399 \\ 
$\expval{v_2} (\%)$ & 7.247 & 7.309 \\
$\expval{v_3}$ (\%) & 2.578 & 2.621 \\
$\expval{v_4}$ (\%) & 1.563 & 1.553 \\
$\expval{v_5}$ (\%) & 1.327 & 1.274 \\
$\expval{[p_T]}$ (MeV) & 631.21 & 629.21 \\
$\rho([p_T], v_2^2)$ & 0.1708 & 0.1773 \\
\hline
\end{tabular}
\end{table}

From these few results, we would conclude that the precise matching prescription used has little impact on ``global'' observables averaged over events. 
More differential observables, in particular those sensitive to event-by-event fluctuations, may be more impacted. 
However, for the 5000 events in the 40--50\% centrality class (see Tab.~\ref{tab:2}) the Pearson coefficient $\rho([p_T], v_2^2)$ changes from 0.1708 with energy matching to 0.1773 with entropy matching, i.e.\ by only a few percent. 
In contrast, the average charged multiplicity now changes by almost 10 percent, although both values remain in good agreement with the ALICE data.
Note that our using the regulators in MUSIC may partly suppress the influence of the different bulk pressures (Fig.~\ref{fig:p_and_Pi}, right panel) in the hydrodynamic stage on final-state observables.

\section{Discussion}
\label{s:discussion}

In this paper, we have tested a new prescription for the switch to a fluid-dynamical description (MUSIC) from a pre-equilibrium model (\Kompost) with a different equation of state.
More precisely, we imposed the continuity in each fluid cell of the equivalent equilibrium entropy, instead of imposing local energy conservation as is usually done. 
We compared two sets of MC Glauber + \Kompost\ + MUSIC + iSS+ SMASH simulations, starting from the same ensemble of initial conditions with different overall normalizations according to the matching prescription used. 
For those simulated events, we found that the difference in matching recipe only has a minor impact on the observables we computed, with the possible exception of the most differential one, namely the correlation between the fluctuations of elliptic flow and average transverse momentum. 
Focusing on the switching time between \Kompost\ and MUSIC, the new entropy-matching prescription leads to a smaller bulk pressure, and also a smaller ratio $\Pi/P$, in the initial state of the fluid-dynamical evolution. 

It is clear that the proposed matching recipe has the unpleasant feature that it does not conserve energy at the interface between pre-equilibrium and fluid dynamics. 
This is admittedly inelegant, but one should remember that in the present state of the art, the absolute energy density at the beginning of fluid dynamics is determined from the multiplicity in the final state of the evolution, from which an ad hoc $K$-factor in the initial state (or sometimes at the beginning of hydrodynamics~\cite{Liu:2015nwa}) is deduced. 
Since multiplicity is rather associated with total entropy, a prescription based on the latter would seem more motivated\footnote{See Ref.~\cite{Huovinen:2003fa} for a discussion of whether energy or entropy density should be used in the initial state of fluid dynamics.} --- at least as long as there is no initial-state model that predicts the value of the initial energy without any indeterminate $K$-factor.
Now, our approach in the present study does not exactly conserve entropy at the switch between \Kompost\ and MUSIC, but it is already a significant step in that direction, since the jump in total entropy is only of second order in the gradients, as discussed at the end of Sec.~\ref{s:setup}. 

The other strong point of our prescription is the significant reduction of the ratio of bulk over thermodynamic pressure in the initial state of fluid dynamics, which means that the applicability of the latter is better warranted. 
This would be even more the case in smaller systems~\cite{Grosse-Oetringhaus:2024bwr}, in which ``the edges'' constitute a larger fraction of the fireball, or if the switching time $\tau_{\rm hydro}$ from pre-equilibrium to hydrodynamics is such that the LQCD equation of state is farther from conformal than what we assumed in this exploratory study.
Accordingly, we plan to further investigate the impact of the entropy-matching recipe, with a more systematic study of final-state observables, in particular sensitive to event-by-event fluctuations, and also across different system sizes, and with various choices for $\tau_{\rm hydro}$.

Eventually, let us emphasize that here we chose \Kompost\ and MUSIC for our simulations, but the proposed entropy-matching recipe may be used (and tested) with any other pre-equilibrium or fluid-dynamical model.

\begin{acknowledgments}

We thank Travis Dore, Gr\'egoire Pihan, S\"oren Schlichting, Chun Shen, and Jie Zhu for helpful discussions and comments. 
N.~B. and R.~K. acknowledge support by the Deutsche Forschungsgemeinschaft (DFG, German Research Foundation) through the CRC-TR 211 'Strong-interaction matter under extreme conditions' - project number 315477589 - TRR 211.
H.~R. was supported by the National Science Foundation (NSF) within the framework of the JETSCAPE collaboration (OAC-2004571) and by the DOE (DE-SC0024232).
Numerical simulations presented in this work were performed at the Paderborn Center for Parallel Computing (PC$^2$) and we gratefully acknowledge their support.

\end{acknowledgments}

\end{document}